\documentclass[aps,11pt,nofootinbib,preprintnumbers,showpacs]{revtex4}

\usepackage{amssymb, amsfonts, amsmath, bm}
\usepackage{graphicx}
\usepackage{color}
\begin{document}

\title{{\Large
Gravitational solitons in Levi-Civit\`a spacetime
}}

\hfill{\small RUP-15-8}

\author{Takahisa Igata$^{1,2}$}
\email{igata@rikkyo.ac.jp}
\author{Shinya Tomizawa$^3$}
\email{tomizawasny@stf.teu.ac.jp}
\affiliation{
${}^1$Department of Physics, Rikkyo University, Toshima, Tokyo 175-8501, Japan\\
${}^2$Graduate School of Science and Technology, 
Kwansei Gakuin University, Sanda, Hyogo 669-1337, Japan\\
${}^3$Department of Liberal Arts, 
Tokyo University of Technology, Otaku, Tokyo 144-8535, Japan
}

%Exact solutions, Gravitational waves

%%%%%%%%%%
\begin{abstract}
Applying the Pomeransky inverse scattering method 
to the four-dimensional vacuum Einstein equation and 
using the Levi-Civit\`a solution for a seed, 
we construct a cylindrically symmetric single-soliton solution. 
Although the Levi-Civit\`a spacetime generally includes 
singularities on its axis of symmetry, 
it is shown that for the obtained single-soliton solution, 
such singularities can be removed by choice of certain special parameters. 
This single-soliton solution describes propagation of 
nonlinear cylindrical gravitational shock wave pulses rather than solitonic waves.  
By analyzing wave amplitudes and time-dependence of polarization angles, 
we provides physical description of the single-soliton solution. 
\end{abstract}
%%%%%%%%%%

%
\maketitle

%%%%%%%%%%
\section{Introduction}
\label{sec:1}
%%%%%%%%%%
Time-depending gravitational soliton solutions in general relativity 
are interpreted as gravitational solitonic waves 
propagating in background spacetimes. 
The so-called inverse scattering method, 
which was established by Belinski and Zakharov, 
has been used as one of the powerful tools 
to construct such soliton solutions~\cite{Belinsky:1971nt, Belinsky:1979mh}. 
In fact, a lot of soliton solutions describing nonlinear gravitational waves 
have been subsequently found 
by using such a systematic method~\cite{Belinski:2001ph, Stephani:2003tm}. 
It is a noteworthy fact that, 
in a stationary and axisymmetric case, 
the application of the method to the four-dimensional vacuum Einstein equation 
can generate exact solutions of black holes. 
However, the simple generalization of the method to higher dimensions 
generally leads to singular solutions. 
Under these circumstances, 
Pomeransky~\cite{Pomeransky:2005sj} modified 
the original inverse scattering method 
so that it could generate regular black hole solutions even in higher dimensions. 
Thereafter, this improved method has played an important role 
in solution-generating of five-dimensional 
black holes solutions~\cite{Iguchi:2011qi, LivingReview}.

Gravitational solitons with cylindrical symmetry 
are of special interest to 
a lot of relativists since they give us 
the simplest treatment of gravitational waves in an exact form. 
While Piran \textit{et al}.~\cite{Piran:1985dk} numerically studied 
nonlinear interactions of cylindrical gravitational waves with two polarization modes, 
Tomimatsu~\cite{Tomimatsu:1989vw} first analytically studied
such nonlinear phenomenon as the gravitational Faraday rotation 
for the cylindrical gravitational solitons generated 
by the Belinski-Zakharav inverse scattering technique. 
Moreover, the interactions of gravitational soliton waves 
with a cosmic string were also discussed 
in Refs.~\cite{Xanthopoulos:1986bx, Xanthopoulos:1986ac, Economou:1988gu, Economou:1988yr}.
Recently, one of the authors~\cite{Tomizawa:2013soa, Tomizawa:2015zva} 
analyzed one- and two-soliton solutions constructed 
by the Pomeransky improved inverse scattering method, 
and studied nonlinear effects of gravitational waves 
such as the gravitational Faraday rotation and time shift phenomenon.

Most soliton solutions generated by the inverse scattering method 
can be obtained by the soliton transformation from seeds 
with a diagonal form. 
In particular, in a cylindrically symmetric case, 
one example of a diagonal metric is the Levi-Civit\`a family, 
which describes the static and cylindrically symmetric spacetime 
labeled by two parameters. 
Therefore this solution can be regarded as the exterior field of 
an infinite cylinder with uniform mass, 
and in general it has naked singularities on its axis of symmetry. 
However, within a certain parameter range, 
this kind of the singularities can be considered as a line source 
with infinite length that yields such cylindrically symmetric gravitational field.

In this paper, 
applying the Pomeransky inverse scattering method 
and using the Levi-Civit\`a metric for a seed, 
we generate a single-soliton solution 
that does not admit staticity but cylindrical symmetry. 
It is a generalization of the solution 
obtained from the Minkowski seed~\cite{Tomizawa:2013soa} 
since the Levi-Civit\`a spacetime includes the Minkowski spacetime 
as a special case. 
Although the Levi-Civit\`a spacetime has singularities 
on the axis except for the Minkowski spacetime, 
for the single-soliton solution, 
such singularities disappear by a certain choice of parameters. 
It is shown that the solution we present in this paper 
describes a shock wave pulse of nonlinear outgoing gravitational waves.

This paper is organized as follows. 
In the following section, 
we construct a single-soliton solution with a real pole
by using the Pomeransky inverse scattering method 
from the Levi-Civit\`a metric as a seed. 
In Sec.~\ref{sec:3}, 
for the single-soliton solution, 
we calculate the amplitudes and polarization angles 
for ingoing and outgoing gravitational waves. 
Moreover, we analyze asymptotic behaviors of 
the nonlinear gravitational waves at the spacetime boundary depending on 
each choice of parameters.
In Sec.~\ref{sec:4}, 
we devote ourselves to the summary and discussion on our results.
Furthermore, we consider the difference from 
the single-soliton solution in Ref.~\cite{Tomizawa:2013soa}, 
which was obtained from the Minkowski seed. 
In Appendix~\ref{sec:A}, 
we review definitions of nonlinear cylindrical gravitational waves 
such as amplitudes and polarization angles, 
which were first introduced 
by Piran \textit{et al.}~\cite{Piran:1985dk} and Tomimatsu~\cite{Tomimatsu:1989vw} .

~~

%%%%%%%%%%
\section{Single-soliton solution}
\label{sec:2}
%%%%%%%%%%
In this section, starting from the Levi-Civit\`a solution, 
we derive a single-soliton solution 
by the Pomeransky inverse scattering method. 
The Levi-Civit\`a solution is a static and cylindrically symmetric solution 
to the four-dimensional vacuum Einstein equation. 
The metric is given in the following form
\begin{align}
ds^2
=\rho^{1+d}\,d\phi^2+\rho^{1-d}\,dz^2+b^2 \rho^{(d^2-1)/2}\left(d\rho^2-dt^2\right),
\label{eq:Levi-Civita}
\end{align}
where $b$ and $d$ are independent parameters, 
and both of them are assumed to be positive without loss of generality. 
For $d=1$, this metric recovers the Minkowski metric 
(with a deficit angle related to $b$) written in the cylindrical coordinates. 
In addition, the symmetry of the Levi-Civit\`a spacetime is enhanced 
due to the existence of an additional Killing vector field 
$\phi\left(\partial/\partial z\right)-z\left(\partial/\partial \phi\right)$ 
for $d=0$ and 
$\phi\left(\partial/\partial t\right)+t\left(\partial/\partial \phi\right)$ 
for $d=3$~\cite{Gautreau:1969qk}. 
It should be noted that this two-parameter family possesses 
naked curvature singularities on its axis of symmetry 
$\rho=0$ except for $d=1$. 
For $d>1$, these singularities can be interpreted 
as a physical gravitational line source 
because a test particle is subjected to an attractive force~\cite{Gautreau:1969qk}. 
In particular, for $d\simeq 1$, 
it can be regarded as the exterior field of 
the infinitely extended cylinder whose mass per unit length is 
\begin{align}
\lambda
=\frac{d-1}{2\left(d+1\right)},
\end{align}
in the Newtonian limit. 
Conversely, for $0\leq d <1$, the singularities on the axis 
cannot be understood as such a physical line source 
because a test particle near its axis suffers 
from the repulsive force by the line source. 
The obtained single-soliton solution, however, 
does not necessarily have a source of repulsive force 
even if the corresponding Levi-Civit\`a seed includes an unphysical source. 
Therefore, in what follows, we use the Levi-Civit\`a metric 
within the range $0\leq d < \infty$ as a seed to generate a single-soliton solution.

Now let us assume that a four-dimensional spacetime admits 
cylindrical symmetry, namely, 
that there exist two commuting Killing vector fields, 
an axisymmetric Killing vector field $\partial/\partial \phi$ 
and a spatially translational Killing vector field $\partial/\partial z$, 
where the polar angle coordinate $\phi$ and the coordinate $z$ have 
the ranges $0\le \phi<\Delta\phi$ and $-\infty<z<\infty$, respectively. 
Under the symmetry assumption, 
the most general metric that is the solution 
to the four-dimensional vacuum Einstein equation
can be described in the Kompaneets-Jordan-Ehlers form: 
\begin{align}
ds^2
=e^{2\psi}\left(dz+\omega\,d\phi\right)^2
+\rho^2\,e^{-2\psi}\,d\phi^2
+e^{2\left(\gamma-\psi\right)}\left(d\rho^2-dt^2\right),
\end{align}
where the functions $\psi$, $\omega$, and $\gamma$ 
depend on the time coordinate $t$ and radial coordinate $\rho$ only. 
Let us define a $2\times 2$ metric $g$ and a metric function $f$ by
\begin{align}
&
g
=\left(
\begin{array}{cc}
e^{2\psi}
&
\omega \,e^{2\psi}
\\
\omega\,e^{2\psi}
&
\rho^2\,e^{-2\psi}+\omega^2\,e^{2\psi}
\end{array}
\right),
\\
&
f=e^{2(\gamma-\psi)}, 
\end{align}
respectively.

For the Levi-Civit\`a metric, 
the $2\times2$ metric $g_0$ and the metric function $f_0$ are written as
\begin{align}
&
g_0
=\mathrm{diag}\Big(\,
\rho^{1-d}, \rho^{1+d}
\,\Big),
\\
&
f_0
=b^2 \rho^{(d^2-1)/2},
\end{align}
respectively.
Following the Pomeransky method~\cite{Pomeransky:2005sj}, 
let us remove a trivial soliton at $t=t_1$ with a trivial BZ vector $(1,0)$, 
and then we have the metric
\begin{align}
g'_{0}
=\mathrm{diag}\left(\rho^{-1-d}\mu^2, \rho^{1+d}
\right)
=
\mathrm{diag}\left(
\frac{\rho^{3-d}}{\tilde{\mu}^2}, \rho^{1+d}
\right),
\label{eq:g0}
\end{align}
where the functions $\mu$ and $\tilde{\mu}$ are defined by
\begin{align}
&
\mu
=\sqrt{(t-t_1)^2-\rho^2}-(t-t_1),
\\
&
\tilde{\mu}
=\frac{\rho^2}{\mu}=-\sqrt{(t-t_1)^2-\rho^2}-(t-t_1),
\end{align}
respectively.

Next, add back a nontrivial soliton with a BZ vector $m_{0}=(1,a)$, 
and then we obtain a single-soliton solution as
\begin{align}
&
g_{ab}
=g'_{0ab}-\frac{g'_{0ac}\,m_c\,\Gamma^{-1}\,m_d\,g'_{0db}}{\mu^2},
\label{eq:1-soliton}
\\
&
f
=f_0\,\frac{\Gamma}{\Gamma_0},
\end{align}
where
\begin{align}
&
\Gamma
=\frac{m_a\,g'_{0ab}\,m_b}{-\rho^2+\mu^2},\\
&
m_a
=m_{0b}\left[\Psi^{-1}_{0}(\mu,\rho,t)\right]_{ba}.
\end{align}
Here, $\Gamma_0$ is $\Gamma$ evaluated at $a=0$, and
$\Psi_0(\lambda, \rho, t)$ is the generating matrix for the metric $g_0'$ in the following form
\begin{align}
\Psi_0(\lambda, \rho, t)
=\mathrm{diag}\left(
\frac{\left(\rho^2+2\,t\, \lambda+\lambda^2\right)^{\left(3-d\right)/2}}{\left(\tilde\mu-\lambda\right)^2},\,
(\rho^2+2\,t\, \lambda+\lambda^2)^{\left(1+d\right)/2}\right),
\end{align}
where $\lambda$ is the spectral parameter. 
Finally, we reparameterize $a$ as $a\left(-2\,t_1\right)^{1-d}\to a$, 
and shift the time coordinate $t$ as $t\to t+t_1$.

Thus, from Eq.~\eqref{eq:1-soliton}, 
we can read off the functions $\psi$, $\gamma$, and $\omega$ 
for the single-soliton solution as
\begin{align}
&
e^{2\psi}=
\frac{1+a^2F\, w^2}{1+a^2F}
\,\rho^{1-d},
\\
&
\omega=
\frac{a\,w^{1-d}}{\left(1-w^2\right)\left(1+a^2 F\,w^2\right)}
\,\rho^{d-1},
\\
&
e^{2\gamma}=
b^2
\left(1+a^2 F\,w^2\right)
\rho^{\left(d-1\right)^2/2},
\end{align}
respectively, with
\begin{align}
&
F
=\frac{w^{2\left(2-d\right)}}{\rho^2\left(1-w^2\right)^4},
\\
&
w=-\frac{\sqrt{t^2-\rho^2}-t}{\rho}.
\end{align}
This single-soliton solution includes three parameters $b$, $d$, and $a$, 
where $b$ and $d$ are assumed to be positive. 
Note that for $a=0$ this metric recovers the Levi-Civit\`a metric. 
Except for $a=0$ the metric depends on the time coordinate $t$ 
as well as the radial coordinate $\rho$. 
Furthermore, for $d=1$, it completely coincides 
with the single-soliton solution in Ref.~\cite{Tomizawa:2013soa}, 
which was generated from the Minkowski seed. 

~~

%%%%%%%%%%
\section{Analysis for the single-soliton solution}
\label{sec:3}
%%%%%%%%%%
In this section, 
we analyze physical properties of 
the nonlinear cylindrically symmetric gravitational waves 
described by the obtained single-soliton solution 
by seeing wave amplitudes, polarization angles, 
and \textit{C}-energy density (see Appendix~\ref{sec:A} for their definitions). 
In particular, we investigate the dependence of 
their asymptotic behaviors on $d$ 
in the neighborhood of its spacetime boundaries.

From the definitions given in Eqs.~\eqref{eq:Ap_def} and \eqref{eq:Bp_def}, 
the amplitudes of ingoing and outgoing waves with the $+$ mode, $A_+$ and $B_+$, 
are calculated, respectively, as
\begin{align}
&
A_+
=\frac{2\,a^2 F}{\rho}\,\sqrt{\frac{u}{v}}\,\,\frac{
1+a^2 F\,w^2-\left(1-w\right)^2+\left(d-1\right)\left(1-w^2\right)}
{\left(1+a^2 F\right)\left(1+a^2 F\,w^2\right)}
-\frac{d-1}{\rho},
\label{eq:Ap}
\\
&
B_+
=-\frac{2\,a^2 F}{\rho}\,\sqrt{\frac{v}{u}}\,\frac{
1+a^2 F\,w^2-\left(1+w\right)^2+\left(d-1\right)\left(1-w^2\right)}
{\left(1+a^2 F\right)\left(1+a^2 F\,w^2\right)}
+\frac{d-1}{\rho},
\label{eq:Bp}
\end{align}
and, from Eqs.~\eqref{eq:Ac_def}--\eqref{eq:Bc_def}, 
the amplitudes of ingoing and outgoing waves with the $\times$ mode, 
$A_\times$ and $B_\times$, are obtained as 
\begin{align}
&
A_\times
=\frac{2\,a\sqrt{F}}{\rho}\,\sqrt{\frac{u}{v}}\,\,
\frac{w\left(1-a^2 F+2\,a^2 F\,w\right)+\left(d-1\right)\left(1+w\right)\left(1+a^2 F\,w\right)}
{\left(1+a^2 F\right)\left(1+a^2 F\,w^2\right)},
\label{eq:Ac}
\\
&
B_\times
=-\frac{2\,a\sqrt{F}}{\rho}\,\sqrt{\frac{v}{u}}\,\,
\frac{w\left(1-a^2 F-2\,a^2 F\,w\right)-\left(d-1\right)\left(1-w\right)\left(1-a^2 F\,w\right)}
{\left(1+a^2 F\right)\left(1+a^2 F\,w^2\right)},
\label{eq:Bc}
\end{align}
respectively. From Eqs.~\eqref{eq:A_def} and \eqref{eq:B_def}, 
the total amplitudes for ingoing and outgoing waves, $A$ and $B$, 
are calculated, respectively, as
\begin{align}
&
A
=\frac{1}{\rho}\,\sqrt{
\frac{a^2 F\,w^2\,\big(\,2\sqrt{u/v}+1-d\,\big)^2+\left(d-1\right)^2}{1+a^2 F\,w^2}
},
\label{eq:A}
\\
&
B
=\frac{1}{\rho}\,\sqrt{
\frac{a^2 F\,w^2\,\big(2\sqrt{v/u}+1-d\,\big)^2+\big(d-1\big)^2}{1+a^2 F\,w^2}
}.
\label{eq:B}
\end{align}
The polarization angles for ingoing and outgoing waves, 
$\theta_A$ and $\theta_B$, given 
in Eqs.~\eqref{eq:thetaA_def} and \eqref{eq:thetaB_def} 
are written as 
\begin{align}
&
\sin 2\theta_A
=2\,a\sqrt{F}\,\sqrt{\frac{u}{v}}\,
\frac{w\left(1-a^2 F+2\,a^2 F\,w\right)+\left(d-1\right)\left(1+w\right)\left(1+a^2 F\,w\right)}
{\left(1+a^2 F\right)\sqrt{1+a^2 F\,w^2}\,\sqrt{a^2 F\,w^2\,\big(\,2\sqrt{u/v}+1-d\,\big)^2+\left(d-1\right)^2}},
\label{eq:thetaA}
\\
&
\sin 2\theta_B
=-2\,a\sqrt{F}\,\sqrt{\frac{v}{u}}\,
\frac{w\left(1-a^2 F-2\,a^2 F\,w\right)-\left(d-1\right)\left(1-w\right)\left(1-a^2 F\,w\right)}
{\left(1+a^2 F\right)\sqrt{1+a^2 F\,w^2}\,\sqrt{a^2 F\,w^2\,\big(2\sqrt{v/u}+1-d\,\big)^2+\big(d-1\big)^2}},
\label{eq:thetaB}
\end{align}
respectively. 
The \textit{C}-energy density is proportional to $\gamma_{,\rho}$, 
which is related to the amplitudes as Eq.~\eqref{eq:C-energy_def}, 
and is given by
\begin{align}
\gamma_{,\rho}
=\frac{\left(d-1\right)^2}{4\,\rho}
+\frac{a^2 F\,w^2}{\rho}\,
\frac{2\left(1+3\,w^2\right)-d\left(1-w^4\right)}{\left(1-w^2\right)^2\left(1+a^2 F\,w^2\right)}.
\label{eq:C-energy}
\end{align}
In what follows, we focus only on the portion $t\geq0$ in the spacetime 
because our interest here is to understand 
how shock wave pulses propagate throughout the spacetime as time passes. 
In the following subsections, 
we analyze the asymptotic behaviors of the above quantities 
near the spacetime boundaries: 
the axis of symmetry $\rho=0$, the light cone $u=0$, 
timelike infinity $t\to\infty$, and null infinity $v\to \infty$. 

~~

%%%%%%%%%%
\subsection{Axis of symmetry}
%%%%%%%%%%

%%%%%%%%%%
\begin{table}[t]
\scalebox{0.8}[0.8]{
\begin{tabular}{cccccccccccccc}
\hline
\hline
$d$
&~~~~~&
$A_+$
&~~~&
$B_+$
&~~~&
$A_\times$
&~~~&
$B_\times$
&~~~&
$A$
&~~~&
$B$&~
\\
\hline
\\[-3mm]
$d=0$
&&
$\dfrac{1}{\rho}$
&&
$-\dfrac{1}{\rho}$
&&
$-\dfrac{a}{2\,t_0^{2}}$
&&
$-\dfrac{a}{2\,t_0^{2}}$
&&
$\dfrac{1}{\rho}$
&&
$\dfrac{1}{\rho}$
&
\\[3mm]
$0< d<1$
&&
$\dfrac{1-d}{\rho}$
&&
$-\dfrac{1-d}{\rho}$
&&
$-\dfrac{2\,a\left(1-d\right)}{\left(2\,t_0\right)^{2-d}\rho^{d}}$
&&
$-\dfrac{2\,a\left(1-d\right)}{\left(2\,t_0\right)^{2-d}\rho^{d}}$
&&
$\dfrac{1-d}{\rho}$
&&
$\dfrac{1-d}{\rho}$
&
\\[3mm]
$d=1$
&&
$\dfrac{2\,a^2}{t_0\left(4\,t_0^{\,2}+a^2\right)}$
&&
$\dfrac{2\,a^2}{t_0\left(4\,t_0^{\,2}+a^2\right)}$
&&
$\dfrac{a}{2\,t_0^{\,2}}\,\dfrac{4\,t_0^{\,2}-a^2}{4\,t_0^{\,2}+a^2}$
&&
$-\dfrac{a}{2\,t_0^{\,2}}\,\dfrac{4\,t_0^{\,2}-a^2}{4\,t_0^{\,2}+a^2}$
&&
$\dfrac{|\,a\,|}{2\,t_0^{\,2}}$
&&
$\dfrac{|\,a\,|}{2\,t_0^{\,2}}$
&
\\[3mm]
$1<d<\dfrac{3}{2}$
&&
$\dfrac{d-1}{\rho}$
&&
$-\dfrac{d-1}{\rho}$
&&
$\dfrac{2\left(d-1\right)\left(2\,t_0\right)^{2-d}}{a\,\rho^{2-d}}$
&&
$\dfrac{2\left(d-1\right)\left(2\,t_0\right)^{2-d}}{a\,\rho^{2-d}}$
&&
$\dfrac{d-1}{\rho}$
&&
$\dfrac{d-1}{\rho}$
&
\\[3mm]
$d=\dfrac{3}{2}$
&&
$\dfrac{1}{2\,\rho}$
&&
$-\dfrac{1}{2\,\rho}$
&&
$\dfrac{4\,t_0^{\,2}-a^2}{2\sqrt{2}\,a\,t_0^{3/2}\sqrt{\rho}}$
&&
$\dfrac{4\,t_0^{\,2}+a^2}{2\sqrt{2}\,a\,t_0^{3/2}\sqrt{\rho}}$
&&
$\dfrac{1}{2\,\rho}$
&&
$\dfrac{1}{2\,\rho}$
&
\\[3mm]
$\dfrac{3}{2}<d<2$
&&
$\dfrac{d-1}{\rho}$
&&
$-\dfrac{d-1}{\rho}$
&&
$-\dfrac{2\,a\left(2-d\right)}{\left(2\,t_0\right)^{3-d}\rho^{d-1}}$
&&
$\dfrac{2\,a\left(2-d\right)}{\left(2\,t_0\right)^{3-d}\rho^{d-1}}$
&&
$\dfrac{d-1}{\rho}$
&&
$\dfrac{d-1}{\rho}$
&
\\[3mm]
$d=2$
&&
$\dfrac{1}{\rho}$
&&
$-\dfrac{1}{\rho}$
&&
$\dfrac{2}{a}+\dfrac{4\,a}{4\,t_0^{\,2}+a^2}$
&&
$\dfrac{2}{a}+\dfrac{4\,a}{4\,t_0^{\,2}+a^2}$
&&
$\dfrac{1}{\rho}$
&&
$\dfrac{1}{\rho}$
&
\\[3mm]
$2<d\ (d\neq 3)$
&&
$\dfrac{3-d}{\rho}$
&&
$-\dfrac{3-d}{\rho}$
&&
$\dfrac{2\left(d-2\right)\left(2\,t_0\right)^{3-d}}{a\,\rho^{3-d}}$
&&
$-\dfrac{2\left(d-2\right)\left(2\,t_0\right)^{3-d}}{a\,\rho^{3-d}}$
&&
$\dfrac{|\,3-d\,|}{\rho}$
&&
$\dfrac{|\,3-d\,|}{\rho}$
&
\\[3mm]
$d=3$
&&
$-\dfrac{2}{t_0}$
&&
$-\dfrac{2}{t_0}$
&&
$\dfrac{2}{a}$
&&
$-\dfrac{2}{a}$
&&
$\dfrac{2}{|\,a\,|}\,\sqrt{1+\dfrac{a^2}{t_0^{\,2}}}$
&&
$\dfrac{2}{|\,a\,|}\,\sqrt{1+\dfrac{a^2}{t_0^{\,2}}}$
&
\\[3mm]
\hline
\hline
\end{tabular}
}%end_\scalebox
\caption{
Asymptotic behaviors of the amplitudes near the axis of symmetry $\rho=0$. }
\label{table:A}
\bigskip
\end{table}
%%%%%%%%%%

Let us see the asymptotic behaviors of wave amplitudes, 
polarization angles, and \textit{C}-energy density near the axis $\rho=0$. 
In the limit $\rho\to 0$ with the time coordinate fixed at $t=t_0$, 
the \textit{C}-energy density \eqref{eq:C-energy} behaves as
\begin{align}
\gamma_{,\rho}
=O(\rho^{-1}),
\end{align}
for all $d$ except for $d=1, 3$. 
Hence, for $d\neq1,3$, this spacetime has a singular gravitational source on the axis.

As shown in Ref.~\cite{Tomizawa:2013soa}, in contrast, for $d=1$, 
it behaves as $\gamma_{,\rho}= O(\rho)$, 
which implies that there is no singular source on the axis. 
This may not be surprising 
because the corresponding seed (Minkowski spacetime) 
has no singular source on the axis. 
However, it should be a surprising fact that 
Eq.~\eqref{eq:C-energy} for $d=3$ asymptotically behaves as 
\begin{align}
&\gamma_{,\rho}
\simeq \frac{t_0^{\,2}+a^2}{a^2\,t_0^{\,2}}\,\rho, 
\end{align}
near $\rho=0$ since, as mentioned in the previous section, 
the Levi-Civit\`a spacetime with $d=3$ 
has singularities on the axis.
As a result, such singularities on the axis is completely 
removed after the soliton transformation. 
Therefore, for $d=3$, nonexistence of curvature singularities on the axis 
allows to evaluate a deficit angle as a meaningful physical quantity. 
The deficit angle $\Delta$ on the axis at arbitrary time is calculated as
\begin{align}
\Delta
&=2\pi-\lim_{\rho\to0}\,
\frac{\displaystyle{\int^{\Delta\phi}_0\sqrt{g_{\phi\phi}}\,d\phi}}
{\displaystyle{\int^\rho_0\sqrt{g_{\rho\rho}}\,d\rho}}
\\
&=2\pi -\frac{\Delta\phi}{b\,|\,a\,|}.
\end{align}
Hence, the deficit angle can be adjusted to be zero 
by choosing the periodicity of $\phi$ as $\Delta\phi=2\pi\,b\,|\,a\,|$.

Table~\ref{table:A} shows the asymptotic behaviors of 
the amplitudes near the axis for each $d$.
While for $0\leq d\le3$ except $d=1,3$, 
the wave amplitudes diverge on the axis, for $d=1,3$ 
all of the amplitudes take finite values there. 
Note that, for $d=0, 2$, the wave amplitudes with the $\times$ mode take finite values, 
and hence the $+$ mode dominates over the $\times$ mode on the axis.
For $d>3$, the wave amplitudes with the $\times$ mode vanish on the axis. 

~~

%%%%%%%%%%
\subsection{Light cone}
%%%%%%%%%%

%%%%%%%%%%
\begin{table}[t]
\scalebox{0.9}[0.9]{
\begin{tabular}{cccccccccccccc}
\hline\hline
$d$
&~~~~~&
$A_+$
&~~~&
$B_+$
&~~~&
$A_\times$
&~~~&
$B_\times$
&~~~&
$A$
&~~~&
$B$
&~
\\
\hline
\\[-3mm]
$d\neq 1,\dfrac12$
&&
$\dfrac{1-d}{v_0}$
&&
$-\dfrac{2}{\sqrt{v_0u}}$
&&
$\dfrac{32\left(2\,d-1\right)u^{3/2}}{a\,v_0^{3/2}}$
&&
$\dfrac{96}{a}\sqrt{\dfrac{u}{v_0}}$
&&
$\dfrac{|\,1-d\,|}{v_0}$
&&
$\dfrac{2}{\sqrt{v_0u}}$
&
\\[3mm]
$d=\dfrac12$
&&
$\dfrac{1}{2\,v_0}$
&&
$-\dfrac{2}{\sqrt{v_0u}}$
&&
$-\dfrac{96\,u^2}{a\,v_0^2}$
&&
$\dfrac{96}{a}\sqrt{\dfrac{u}{v_0}}$
&&
$\dfrac{1}{2\,v_0}$
&&
$\dfrac{2}{\sqrt{v_0u}}$
&
\\[3mm]
$d=1$
&&
$\dfrac{2\,\sqrt{u}}{v_0^{3/2}}$
&&
$-\dfrac{2}{\sqrt{v_0u}}$
&&
$\dfrac{32\,u^{3/2}}{a\,v_0^{3/2}}$
&&
$\dfrac{96}{a}\sqrt{\dfrac{u}{v_0}}$
&&
$\dfrac{2\,\sqrt{u}}{v_0^{3/2}}$
&&
$\dfrac{2}{\sqrt{v_0u}}$
&
\\[3mm]
\hline
\hline
\end{tabular}
}
\caption{
Asymptotic behaviors of the amplitudes near the light cone $u=0$. }
\label{table:B}
\bigskip
\end{table}
%%%%%%%%%%

We turn our attention to the asymptotic behaviors 
of the gravitational waves near the light cone $t=\rho\ (u=0)$ or, equivalently, $w=1$.
Regardless of $d$, the \textit{C}-energy density \eqref{eq:C-energy} diverges there as
\begin{align}
\gamma_{,\rho} 
\simeq 
\frac{2}{\rho\left(1-w\right)^2}.
\end{align}
The divergence of the \textit{C}-energy density leads to curvature singularities, 
whose appearance is commonly unavoidable for the single-soliton solutions. 
Hence, the spacetime region cannot be 
analytically extended to the exterior region over $w=1$. 
We can interpret the (curvature) singularities on the light cone 
as the gravitational shock wave propagating 
at the light velocity from the axis $\rho=0$ at the moment of $t=0$ 
to null infinity $v\to\infty$.

As shown in Table~\ref{table:B}, 
since the amplitudes of the $\times$ mode waves asymptotically approaches to zero 
on the light cone, the $\times$ mode do not contribute to the shock waves. 
The main ingredient of the shock waves is made from the outgoing wave with 
the $+$ mode, whose amplitude diverges there. 
While for $d=1$, no ingoing wave crosses the light cone, 
for $d\not=1$, the ingoing waves with the $+$ mode exist on the light cone. 

~~

%%%%%%%%%%
\subsection{Timelike infinity}
%%%%%%%%%%

%%%%%%%%%%
\begin{table}[t]
\scalebox{0.8}[0.8]{
\begin{tabular}{cccccccccccccc}
\hline\hline
$d$
&~~~~~&
$A_+$
&~~~&
$B_+$
&~~~&
$A_\times$
&~~~&
$B_\times$
&~~~&
$A$
&~~~&
$B$
&~
\\
\hline
\\[-3mm]
$0\leq d <1$
&&
$\dfrac{1-d}{\rho_0}$
&&
$-\dfrac{1-d}{\rho_0}$
&&
$-\dfrac{2\,a\left(1-d\right)}{\rho_0^{\,d}\left(2\,t\right)^{2-d}}$
&&
$-\dfrac{2\,a\left(1-d\right)}{\rho_0^{\,d}\left(2\,t\right)^{2-d}}$
&&
$\dfrac{1-d}{\rho_0}$
&&
$\dfrac{1-d}{\rho_0}$
&
\\[3mm]
$d=1$
&&
$\dfrac{a^2}{2\,t^3}$
&&
$\dfrac{a^2}{2\,t^3}$
&&
$\dfrac{a}{2\,t^2}$
&&
$-\dfrac{a}{2\,t^2}$
&&
$\dfrac{|\,a\,|}{2\,t^2}$
&&
$\dfrac{|\,a\,|}{2\,t^2}$
&
\\[3mm]
$1< d <2$
&&
$\dfrac{1-d}{\rho_0}$
&&
$-\dfrac{1-d}{\rho_0}$
&&
$\dfrac{2\,a\left(d-1\right)}{\rho_0^{\,d}\left(2\,t\right)^{2-d}}$
&&
$\dfrac{2\,a\left(d-1\right)}{\rho_0^{\,d}\left(2\,t\right)^{2-d}}$
&&
$\dfrac{d-1}{\rho_0}$
&&
$\dfrac{d-1}{\rho_0}$
&
\\[3mm]
$d=2$
&&
$\dfrac{1}{\rho_0}-\dfrac{2\,\rho_0}{\rho_0^{\,2}+a^2}$
&&
$-\dfrac{1}{\rho_0}+\dfrac{2\,\rho_0}{\rho_0^{\,2}+a^2}$
&&
$\dfrac{2\,a}{\rho_0^{\,2}+a^2}$
&&
$\dfrac{2\,a}{\rho_0^{\,2}+a^2}$
&&
$\dfrac{1}{\rho_0}$
&&
$\dfrac{1}{\rho_0}$
&
\\[3mm]
$2< d <5/2$
&&
$\dfrac{d-1}{\rho_0}$
&&
$-\dfrac{d-1}{\rho_0}$
&&
$\dfrac{2\left(d-1\right)}{a\,\rho_0^{2-d}\,\left(2\,t\right)^{d-2}}$
&&
$\dfrac{2\left(d-1\right)}{a\,\rho_0^{2-d}\,\left(2\,t\right)^{d-2}}$
&&
$\dfrac{d-1}{\rho_0}$
&&
$\dfrac{d-1}{\rho_0}$
&
\\[3mm]
$d=5/2$
&&
$\dfrac{3}{2\,\rho_0}$
&&
$-\dfrac{3}{2\,\rho_0}$
&&
$\dfrac{3\,\rho_0^{\,2}+a^2}{a\,\rho_0^{3/2}\sqrt{2\,t}}$
&&
$\dfrac{3\,\rho_0^{\,2}-a^2}{a\,\rho_0^{3/2}\sqrt{2\,t}}$
&&
$\dfrac{3}{2\,\rho_0}$
&&
$\dfrac{3}{2\,\rho_0}$
&
\\[3mm]
$\dfrac{5}{2}<d<3$
&&
$\dfrac{d-1}{\rho_0}$
&&
$-\dfrac{d-1}{\rho_0}$
&&
$\dfrac{2\,a\left(d-2\right)}{\rho_0^{d-1}\left(2\,t\right)^{3-d}}$
&&
$-\dfrac{2\,a\left(d-2\right)}{\rho_0^{d-1}\left(2\,t\right)^{3-d}}$
&&
$\dfrac{d-1}{\rho_0}$
&&
$\dfrac{d-1}{\rho_0}$
&
\\[3mm]
$d=3$
&&
$\dfrac{2\,\rho_0}{\rho_0^{\,2}+a^2}$
&&
$-\dfrac{2\,\rho_0}{\rho_0^{\,2}+a^2}$
&&
$\dfrac{2\,a}{\rho_0^{\,2}+a^2}$
&&
$-\dfrac{2\,a}{\rho_0^{\,2}+a^2}$
&&
$ \dfrac{2}{\sqrt{\rho_0^{\,2}+a^2}}$
&&
$ \dfrac{2}{\sqrt{\rho_0^{\,2}+a^2}}$
&
\\[3mm]
$3<d$
&&
$-\dfrac{d-3}{\rho_0}$
&&
$\dfrac{d-3}{\rho_0}$
&&
$\dfrac{2\left(d-2\right)\,\rho_0^{d-3}}{a\left(2\,t\right)^{d-3}}$
&&
$-\dfrac{2\left(d-2\right)\,\rho_0^{d-3}}{a\left(2\,t\right)^{d-3}}$
&&
$\dfrac{d-3}{\rho_0}$
&&
$\dfrac{d-3}{\rho_0}$
&
\\[3mm]
\hline
\hline
\end{tabular}
}
\caption{
Asymptotic behaviors of the amplitudes at timelike infinity $t\to\infty$. }
\label{table:C}
\bigskip
\end{table}
%%%%%%%%%%

Let us consider the asymptotic behaviors of the gravitational waves
at timelike infinity.
Table \ref{table:C} shows the asymptotic behaviors 
of the amplitudes \eqref{eq:Ap}--\eqref{eq:B} at timelike infinity $t\to \infty$ 
with the radial coordinate $\rho$ kept constant as $\rho=\rho_0>0$. 
As shown in Ref.~\cite{Tomizawa:2013soa}, for $d=1$, 
the spacetime is asymptotically flat at $t\to\infty$, 
and hence both amplitudes $A$ and $B$ vanish. 
Since the polarization angles $\theta_A$ and $\theta_B$ behave as 
$\theta_A=-\theta_B\simeq \pi/4$, 
the $\times$ mode for the ingoing and outgoing waves becomes 
dominant at late time. 
For $d=2$, both amplitudes $A$ and $B$ 
asymptotically approach a nonzero constant 
at $t\to\infty$, 
and the angles $\theta_A$ and $\theta_B$ behave as
\begin{align}
\sin 2\theta_A
\simeq
\sin 2\theta_B
\simeq
\frac{2\,a\,\rho_0}{\rho_0^{\,2}+a^2}.
\end{align}
For $d=3$, both amplitudes $A$ and $B$ become constant at $t\to\infty$, 
and the polarization angles behave as
\begin{align}
\sin 2\theta_A
\simeq 
-\sin 2\theta_B
\simeq 
\frac{a}{\sqrt{\rho_0^{\,2}+a^2}}.
\end{align}
For $d\neq1,2,3$, 
the amplitudes $A$ and $B$ also become constant at $t\to\infty$, 
and the polarization angles $\theta_A$ and $\theta_B$ vanish. 
This means that the $+$ mode for ingoing and outgoing waves 
dominate over the $\times$ mode at late time, 
whose asymptotic behaviors are considerably similar to those 
that the Tomimatsu solution~\cite{Tomimatsu:1989vw} shows.

~~

%%%%%%%%%%
\subsection{Null infinity}
%%%%%%%%%%

%%%%%%%%%%
\begin{table}[t]
\scalebox{0.65}[0.65]{
\begin{tabular}{cccccccccccccc}
\hline\hline
$d$
&~~~~~&
$A_+$
&~~~&
$B_+$
&~~~&
$A_\times$
&~~~&
$B_\times$
&~~~&
$A$
&~~~&
$B$
&~
\\
\hline
\\[-3mm]
$d\neq 1,\dfrac12$
&&
$\dfrac{1-d}{v}$
&&
$-\dfrac{2\,a^2\left(a^2-3\cdot16^2u_0^2\right)}{\left(a^2+16^2u_0^2\right)^2\sqrt{u_0\,v}}$
&&
$\dfrac{32\,a\left(2\,d-1\right)u_0^{3/2}}{\left(a^2+16^2u_0^2\right)v^{3/2}}$
&&
$\dfrac{32\,a\left(3\,a^2-16^2u_0^2\right)}{\left(a^2+16^2u_0^2\right)^2}\sqrt{\dfrac{u_0}{v}}$
&&
$\dfrac{|\,d-1\,|}{v}$
&&
$\dfrac{2\,|\,a\,|}{\sqrt{\left(a^2+16^2u_0^2\right)u_0v}}$
&
\\[3mm]
$d=\dfrac{1}{2}$
&&$
\dfrac{1}{2\,v}$
&&
$-\dfrac{2\,a^2\left(a^2-3\cdot16^2u_0^2\right)}{\left(a^2+16^2u_0^2\right)^2\sqrt{u_0\,v}}$
&&
$-\dfrac{32\,a\,u_0^2\left(3\,a^2+16^2u_0^2\right)}{\left(a^2+16^2u_0^2\right)^2\,v^{2}}$
&&
$\dfrac{32\,a\left(3\,a^2-16^2u_0^2\right)}{\left(a^2+16^2u_0^2\right)^2}\sqrt{\dfrac{u_0}{v}}$
&&
$\dfrac{1}{2\,v}$
&&
$\dfrac{2\,|\,a\,|}{\sqrt{\left(a^2+16^2u_0^2\right)u_0v}}$
&
\\[3mm]
$d=1$
&&
$\dfrac{2\,a^2\sqrt{u_0}}{\left(a^2+16^2u_0^2\right)v^{3/2}}$
&&
$-\dfrac{2\,a^2\left(a^2-3\cdot16^2u_0^2\right)}{\left(a^2+16^2u_0^2\right)^2\sqrt{u_0\,v}}$
&&
$\dfrac{32\,a\,u_0^{3/2}}{\left(a^2+16^2u_0^2\right)v^{3/2}}$
&&
$\dfrac{32\,a\left(3\,a^2-16^2u_0^2\right)}{\left(a^2+16^2u_0^2\right)^2}\sqrt{\dfrac{u_0}{v}}$
&&
$\dfrac{2\,|\,a\,|\,\sqrt{u_0}}{\sqrt{a^2+16^2u_0^2}\,v^{3/2}}$
&&
$\dfrac{2\,|\,a\,|}{\sqrt{\left(a^2+16^2u_0^2\right)u_0v}}$
&
\\[3mm]
\hline
\hline
\end{tabular}
}
\caption{
Asymptotic behaviors of the amplitudes at null infinity $v\to\infty$. }
\label{table:D}
\bigskip
\end{table}
%%%%%%%%%%

Let us focus on the asymptotic behaviors of the gravitational waves 
at null infinity. 
Table~\ref{table:D} shows the asymptotic behaviors of 
Eqs.~\eqref{eq:Ap}--\eqref{eq:B} at null infinity $v\to \infty$ 
as $u=u_0$ ($u_0$: a positive constant). 
As discussed in Ref.~\cite{Tomizawa:2013soa}, 
for $d=1$, 
because the spacetime is asymptotically flat at $v\to\infty$, 
both amplitudes $A$ and $B$ vanish. 
Then the polarization angles approach constant values.\footnote{
For $d=1$, the polarization angles at null infinity behaves as
\begin{align}
&
\sin 2\theta_A
\simeq 
\mathrm{sgn}(a)\,\frac{16\,u_0}{\sqrt{a^2+16^2u_0^2}},
\\
&
\sin2\theta_B
\simeq 
\mathrm{sgn}(a)\,\frac{16\,u_0\left(3\,a^2-16^2u_0^2\right)}{\left(a^2+16^2u_0^2\right)^{3/2}}. 
\end{align}}
For $d=1/2$, the amplitudes $A$ and $B$ go to zero as $v\to \infty$ 
as shown in Table~\ref{table:D}, and then the polarization angles behave as 
\begin{align}
&\sin 2\theta_A
\simeq
-\frac{64\,a\,u_0^2\left(3\,a^2+16^2u_0^2\right)}{\left(a^2+16^2u_0^2\right)^2v},
\\
&\sin2\theta_B
\simeq
\mathrm{sgn}(a)\,\frac{16\,u_0\left(3\,a^2-16^2u_0^2\right)}{\left(a^2+16^2u_0^2\right)^{3/2}}. 
\end{align}
While for ingoing waves, the $+$ mode dominate over the 
$\times$ mode at null infinity since $\theta_A\to 0$, 
for outgoing waves, the polarization angle $\theta_B$ approaches 
constant as $v\to \infty$. 
For $u_0=\sqrt{3}\,|\,a\,|/16$, in particular, $\theta_B$ asymptotically vanishes. 

For $d\not=1,1/2$, 
both amplitudes $A$ and $B$ also vanish at null infinity (see Table~\ref{table:D}). 
From Eqs.~\eqref{eq:thetaA} and \eqref{eq:thetaB}, 
we obtain the asymptotic form of the polarization angles as
\begin{align}
&
\sin 2\theta_A 
\simeq 
\frac{32\left(2\,d-1\right)a\,u_0^{3/2}}{|\,d-1\,|\left(a^2+16^2u_0^2\right)^{3/2}\sqrt{v}},
\\
&
\sin 2\theta_B 
\simeq 
\mathrm{sgn}(a)\,\frac{16u_0 \left(3\,a^2-16^2u_0\right)}{\left(a^2+16^2u_0^2\right)^{3/2}}. 
\end{align}
Similarly to the case of $d=1/2$, the polarization angle $\theta_A$ vanishes at null infinity while 
$\theta_B$ approaches to a constant value.

Thus we find that 
independently of $d$, $\tan \theta_B$ 
asymptotically approaches to 
\begin{align}
\tan 2\theta_B
\simeq 
-\frac{16\,u_0\left(3\,a^2-16^2u_0^2\right)}{a^2-3\cdot 16^2u_0^2}.
\end{align}
Therefore, for $v\to\infty$ with $u_0=\sqrt{3}\,|\,a\,|/16$, 
there only exists $+$ mode outgoing waves, 
while for $v\to\infty$ with $u_0=|\,a\,|\slash(16\sqrt{3})$, 
there only exists $\times$ mode outgoing waves.

~~

%%%%%%%%%%
\section{Summary and Discussion}
\label{sec:4}
%%%%%%%%%%
In this paper, 
applying the Pomeransky inverse scattering method 
to the four-dimensional vacuum Einstein equation, 
we have constructed the cylindrically symmetric 
single-soliton solution with a real pole from the the Levi-Civit\`a seed metric. 
The solution obtained in this work 
has three independent parameters, 
where two of them, $b$ and $d$, are originated from the seed 
and the remaining one $a$ is the BZ-parameter 
appearing in the soliton transformation, 
and the Levi-Civit\`a metric is recovered by setting $a=0$. 
The single-soliton spacetime is interpreted as 
propagation of nonlinear gravitational shock waves with cylindrical symmetry. 
In order to understand physical properties of these gravitational waves, 
we have classified and analyzed the wave amplitudes and polarization angles 
for each value of $d\geq 0$. 

\medskip
We summarize the behaviors of gravitational waves near the spacetime boundaries:
%%%%%%%%%%
\begin{itemize}
\item[(i)] Axis of symmetry ($\rho=0$): \\
\ \ 
Except for the Minkowski spacetime corresponding to $d=1$, 
the Levi-Civit\`a seed metric has singularities on the axis of symmetry $\rho=0$. 
In the same way, for the single-soliton solution with $d\neq1,3$, 
the \textit{C}-energy density diverges on the axis 
due to the existence of such singularities, 
and then the wave amplitudes $A$ and $B$ becomes infinitely large. 
For $d=1,3$, 
however, 
the singularities disappears so long as $a\neq 0$.
For $d=3$, 
the polarization angles of ingoing and outgoing waves 
have finite and nonzero values on the axis, 
and approach $\pi/4$ as time passes. 
\item[(ii)] Light cone ($u=0$): \\
\ \ 
Regardless of $d$, 
the outgoing wave amplitude becomes 
infinitely large on the null surface $u=0$. 
The spacetime has null curvature singularities, 
and the \textit{C}-energy diverges there. 
This itself is not special but common to all known single-soliton solutions 
with cylindrical symmetry. 
The polarization angles for ingoing and outgoing waves vanishes on the surface. 
Thus, we can find that an outgoing shock wave pulse 
with the $+$ mode is initially emitted from the origin of the spacetime. 
\item[(iii)]Timelike infinity ($t\to\infty$):\\
\ \ 
At $t\to\infty$ (with $\rho$ constant), the spacetime 
described by the obtained single-soliton solution 
does not asymptotically approaches to the Minkowski spacetime 
except for $d=1$~\cite{Tomizawa:2013soa}, 
in which case simultaneously both ingoing and outgoing gravitational waves decay. 
While for $d=1$ the $\times$ mode for the ingoing and outgoing waves 
becomes dominant at late time, 
for $d=2,3$ the $+$ and $\times$ modes of the ingoing and outgoing waves 
have a comparable order, 
and for $d\not=1,2,3$ the $+$ mode for both waves comes to be dominant at late time. 
\item[(iv)] Null Infinity ($v\to \infty$ with $u=u_0>0$): \\
\ \ 
Regardless of $d$, 
the amplitudes for ingoing and outgoing waves 
decay at null infinity $v\to\infty$ with $u=u_0$ 
($u_0$ : a positive constant). 
The polarization angle for ingoing waves takes a constant value, 
which depends on $d$ at null infinity. 
In contrast, the one for outgoing waves commonly approaches 
a constant value, independently of $d$. 
\end{itemize}

\medskip
It is well known that 
although the appearance of singularities on the light cone is commonly unavoidable 
for the single-soliton solutions with a real pole, 
such a problem can be resolved for two-soliton solutions 
with two complex conjugate poles 
(for instance, see Refs.~\cite{Tomimatsu:1989vw, Tomizawa:2015zva}).
Therefore, for $d=3$, 
it may be interesting to construct such a two-soliton solution with complex conjugate poles 
because it is expected to be entirely regular everywhere. 
This issue deserves further study.

~~

%%%%%%%%%%
\acknowledgments
%%%%%%%%%%
We would like to thank Takashi Mishima and Tomohiro Harada for 
discussions concerning regularity. 
We also would like to thank Ken-ichi Nakao for useful comments. 
This work was partially supported by a Research Grant 
from the Tokyo Institute of Technology Foundation (T.I.) 
and the Grant-in-Aid for Young Scientists (B) (No.~26800120) 
from Japan Society for the Promotion of Science (S.T.).

~~

\appendix
%%%%%%%%%%
\section{Formulas}
\label{sec:A}
%%%%%%%%%%
In this section, we give definitions of 
amplitudes and polarization angles of 
nonlinear cylindrically symmetric gravitational waves. 
Following Ref.~\cite{Piran:1985dk, Tomimatsu:1989vw}, 
we introduce their amplitudes as
\begin{align}
&
A_+
=2\psi_{,v},
\label{eq:Ap_def}
\\
&
B_+
=2\psi_{,u},
\label{eq:Bp_def}
\\
&
A_\times
=\frac{e^{2\psi}\omega_{,v}}{\rho},
\label{eq:Ac_def}
\\
&
B_\times
=\frac{e^{2\psi}\omega_{,u}}{\rho},
\label{eq:Bc_def}
\end{align}
where $A_+$ and $B_+$ describe ingoing and outgoing waves in the $+$ mode, respectively, 
and $A_\times$ and $B_\times$ denote ingoing and outgoing waves in the $\times$ mode, 
respectively. Now the advanced ingoing and outgoing null coordinates $u$ and $v$ 
are defined by $u=(t-\rho)/2$ and $v=(t+\rho)/2$, respectively. 
Total amplitudes of ingoing and outgoing waves are defined by
\begin{align}
&
A
=\sqrt{A_+^2+A_\times^2},
\label{eq:A_def}
\\
&
B
=\sqrt{B_+^2+B_\times^2},
\label{eq:B_def}
\end{align}
respectively, and polarization angles $\theta_A$ and $\theta_B$ 
for the respective wave amplitudes are given by
\begin{align}
&
\sin2\theta_A
=\frac{A_\times}{A},
\label{eq:thetaA_def}
\\
&
\sin2\theta_B
=\frac{B_\times}{B}.
\label{eq:thetaB_def}
\end{align}

Thus, the vacuum Einstein equation can be written in terms of these quantities. 
Actually, the nonlinear differential equations for the functions 
$\psi$ and $\omega$ are replaced by
\begin{align}
&
A_{+,u}
=\frac{A_+-B_+}{2\,\rho}+A_\times B_\times,
\\
&
B_{+,v}
=\frac{A_+-B_+}{2\,\rho}+A_\times B_\times,
\\
&
A_{\times,u}
=\frac{A_\times+B_\times}{2\,\rho}-A_+ B_\times,
\\
&
B_{\times,v}
=-\frac{A_\times+B_\times}{2\,\rho}-A_\times B_+,
\end{align}
and the function $\gamma$ is determined by
\begin{align}
&
\gamma_{,\rho}
=\frac{\rho}{8}\left(A^2+B^2\right),
\label{eq:C-energy_def}
\\
&
\gamma_{,t}
=\frac{\rho}{8}\left(A^2-B^2\right).
\end{align}

%%%%%%%%%%

%%%%%%%%%%

\end{document}